\documentclass[preprint,aps,superscriptaddress]{revtex4}

\usepackage{graphicx}
\usepackage{dcolumn}
\usepackage{bm}

\begin{document}


\title{Abrupt enhancement of non-centrosymmetry and appearance of the spin-triplet superconducting state  in  Li$_{2}$(Pd$_{1-x}$Pt$_{x}$)$_3$B beyond $x$=0.8}

\author{S. Harada}
\affiliation{Department of Physics, Okayama University, Okayama 700-8530, Japan}
\author{J.J. Zhou}
\affiliation{Institute of Physics and Beijing National Laboratory for Condensed Matter Physics, Chinese Academy of Sciences, Beijing 100190, China}
\author{Y.G. Yao}
\email{ygyao@bit.edu.cn}
\affiliation{School of Physics, Beijing Institute of Technology, Beijing 100081, China}
\affiliation{Institute of Physics and Beijing National Laboratory for Condensed Matter Physics, Chinese Academy of Sciences, Beijing 100190, China}
\author{Y. Inada}%
\affiliation{Department of Science Education, Faculty of Education, Okayama University, Okayama 700-8530, Japan}%
\author{Guo-qing Zheng}%
\email{zheng@psun.phys.okayama-u.ac.jp}
\affiliation{Department of Physics, Okayama University, Okayama 700-8530, Japan}
\affiliation{Institute of Physics and Beijing National Laboratory for Condensed Matter Physics, Chinese Academy of Sciences, Beijing 100190, China}

\date{\today}

\begin{abstract}
We report  synthesis,  $^{195}$Pt, $^{11}$B and $^{7}$Li  NMR measurements, and first-principle band calculation for non-centrosymmetric superconductors Li$_{2}$(Pd$_{1-x}$Pt$_{x}$)$_{3}$B ($x$=0, 0.2, 0.5, 0.8, 0.84, 0.9 and 1). For $0 \leq x \leq 0.8$, the spin-lattice relaxation rate ${1/T_1}$  shows a clear coherence peak just below $T_c$, decreasing exponentially at low temperature,  and the  Knight shift $^{195}K$ decreases below  $T_{\rm c}$. For $x$=0.9 and  1.0, in contrast, ${1/T_1}$ shows no coherence peak but a 
$T^{3}$ variation and $^{195}K$ remains unchanged across $T_{\rm c}$. These results indicate that the  superconducting state  changes drastically from a spin-singlet dominant to a spin-triplet dominant state at $x$=0.8. We find that the distortion of B(Pt,Pd)$_6$  increases abruptly above $x$=0.8, which leads to an abrupt  enhancement of the asymmetric spin-orbit coupling as confirmed by band calculation.  Such  local structure distortion that enhances the extent of inversion-symmetry breaking is primarily responsible for the pairing symmetry evolution.   The insight obtained here provides a new guideline for searching new NCS superconductors with large  spin-triplet component.
\end{abstract}

\maketitle
The  superconducting state in materials without spatial inversion symmetry has been a hot  topic in condensed matter physics.
In superconductors with inversion symmetry, the Cooper pairs must be either in the spin-singlet state or in the spin-triplet state because of parity conservation law. In contrast, in superconductors without inversion symmetry, mixing of the two states is permitted \cite{gorkov, frigeri1, samokhin}. The mixing is determined by the spin-orbit coupling (SOC), which is a subject under  active scrutiny in various sub-fields of physics. 

Non-centrosymmetric (NCS) superconductors provide a possible new route to spin-triplet superconductivity with high $T_c$ above 3 K \cite{nishiyama2}. Furthermore, they may exhibit novel phenomena such as  unusual magnetoelectric effects \cite{Fujimoto} or a Fulde-Ferrel-Larkin-Ovchinikov (FFLO) state with spatially-varying pairing functions \cite{Agterberg}. Recently, NCS superconductors have received renewed interests. It has been proposed that they can show topological quantum phenomena such as edge states and non-Abelian statistics \cite{Sato,Tanaka},  in analogy with  topological insulators \cite{SCZhang,Hasan}.  



 Among various NCS compounds, Li$_2$Pt$_3$B ($T_{\rm c}\sim$2.7 K) \cite{Badica} is of particular interest. Both Li$_2$Pd$_3$B ($T_{\rm c}\sim$7 K) \cite{togano} and Li$_2$Pt$_3$B   crystallize  in a perovskite-like cubic structure composed of distorted octahedral units of B(Pd,Pt)$_6$  and are categorized as $P4{_3}32$ (No. 212) in space group \cite{eibenstein}. There is no inversion center in all directions. Many experiments \cite{nishiyama1,nishiyama2,yuan,takeya},  except  $\mu$sr \cite{Hafliger}, have found that Li$_2$Pt$_3$B  and  Li$_2$Pd$_3$B show quite striking different properties.
NMR \cite{nishiyama1,nishiyama2}, penetration depth  and specific heat measurements \cite{yuan,takeya}  have found  that  there exist nodes in the gap function in Li$_2$Pt$_3$B  while Li$_2$Pd$_3$B is a BCS superconductor. Especially, the NMR results  suggested that the spin-triplet state is dominant in Li$_2$Pt$_3$B \cite{nishiyama2}. Early interpretation  was  that the striking difference of the superconducting properties between isostructual Li$_2$Pt$_3$B and Li$_2$Pd$_3$B is due to   the different strength of the SOC originated from different atomic number ($Z$) of Pt and Pd \cite{nishiyama2,yuan}.  
The atomic SOC is in proportion to $Z^2$, 
which differs by a factor of three between Pt and Pd.

Therefore, in the subsequent  search for new NCS superconductors, great effort was made to include heavy elements, which results in discovery of 
Mg$_{10}$Ir$_{19}$B$_{16}$ \cite{klimczuk}, Ir$_2$Ga$_9$ \cite{Nohara}, Re$_3$W \cite{Wen}, BaPtSi$_3$ \cite{Bauer2}, and Ca(Ir,Pt)Si$_3$ \cite{Eguchi}. However, conventional isotropic superconducting gap was found  in these compounds \cite{tahara,Nohara,Eguchi}, and spin-singlet state was evidenced in Mg$_{10}$Ir$_{19}$B$_{16}$ \cite{tahara}.  
This suggests that  $Z$ is not 
the only parameter determining the singlet-triplet mixing, and other factors await to be revealed. 

 In this work, we synthesized Li$_{2}$(Pd$_{1-x}$Pt$_{x}$)$_3$B ($x$=0.2, 0.5, 0.8, 0.84, 0.9 and 1) 
and studied the evolution of the superconducting state by $^{11}$B,  $^{195}$Pt and $^{7}$Li NMR measurements. We have also re-measured $^{11}$B for the previous $x$=0 sample \cite{nishiyama1} at a lower magnetic field $H$=0.26 T. In contrast to a naive expectation, we find that the evolution in the pairing symmetry in   Li$_{2}$(Pd$_{1-x}$Pt$_{x}$)$_3$B is not continuous but abrupt.  For  $0 \leq x \leq 0.8$, the superconducting state is dominantly spin-singlet state. However, for $0.9 \leq x$, spin-triplet state becomes dominant. The $x$=0.84 compound displays both spin-singlet and spin-triplet characteristics. We find  that
the local structure changes abruptly for $x>$0.8 as to increase the extent of inversion-symmetry breaking, which is an effective factor to enhance the strength of SOC as confirmed by our first-principle band calculation. The abrupt increase of the SOC is responsible for the abrupt change of the superconducting properties. The insight obtained here 
provides a new guideline for searching new NCS superconductors with large  spin-triplet component.
 
\begin{figure}[htbp]
  \begin{center}
\includegraphics[width=75mm]{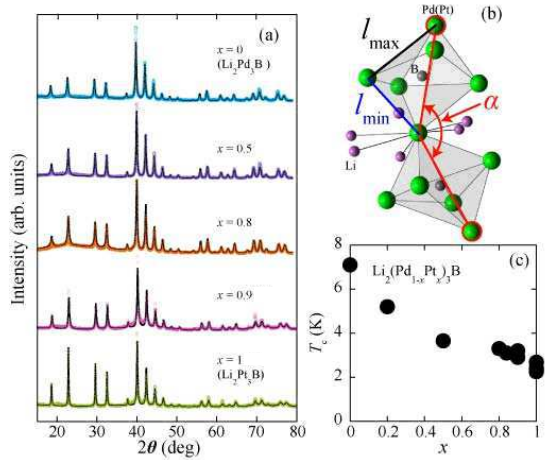}
\end{center}
  \caption{(Color online) (a) Representative Cu-K$_{\alpha}$ XRD charts (circles) with Rietveld analysis (solid curves). 
The error of the Rietveld fitting ($R_{\rm F}$) \cite{izumi} is about 3\%. (b) The distorted B(Pd,Pt)$_6$ octahedral units. $\alpha$ is the angle between the two connected octahedra, and $l_{\rm max}$ ($l_{\rm min}$) is the longest (shortest) M-M (M=Pd,Pt) bond length. (c)    $T_{\rm c}$ versus Pt-content $x$. 
}
\label{xrd}
\end{figure}

 
 Polycrystalline samples of Li$_{2}$(Pd$_{1-x}$Pt$_{x}$)$_3$B ($x$=0.2, 0.5, 0.8, 0.84, 0.9 and 1) were prepared in this study by two-step arc melting method. In the first step, (Pd$_{1-x}$Pt$_{x}$)$_3$B was synthesized by using Pd(99.95\%), Pt(99.999\%) and B(99.8\%).
In the second step, excess Li(99\%) by 5-30\% was added to (Pd$_{1-x}$Pt$_{x}$)$_3$B. For $x$=1, two new samples \#B ($T_{\rm c}$=2.4 K) and \#C ($T_{\rm c}$=2.25 K), whose $T_{\rm c}$ is close to that reported by Yuan {\it et al} ($T_{\rm c}$=2.43 K) \cite{yuan} and Takeya {\it et al} ($T_{\rm c}$=2.17 K) \cite{takeya},  were made and measured. The previously reported $x$=1 sample ($T_{\rm c}$=2.68 K) \cite{nishiyama2} is referred as  $x$=1 \#A. 
All fresh samples were confirmed to be single phase 
by powder X-ray diffraction (XRD) (Fig. \ref{xrd} (a)).  Figure \ref{xrd} (b) depicts the B(Pd,Pt)$_6$ octahedral units whose bond length and the angle $\alpha$ is obtained from Rietveld analysis \cite{izumi}. 
Inductively Coupled Plasma (ICP) analysis was applied to check  the  Li:Pd(Pt):B ratio of the resultant samples \cite{Supple}.  
$T_{\rm c}$ for each sample at $H$=0 and a finite $H$ was determined by measuring the inductance of the {\it in situ} NMR coil. The $T_{\rm c}$ showed a smooth decrease with increasing $x$ (Fig. \ref{xrd} (c)).

NMR measurements were conducted  at   $H$=0.26 T in order to minimize the reduction of $T_{\rm c}$ by $H$. The NMR spectra were obtained by fast Fourier transform  of the spin echo taken at the fixed $H$. The spin-lattice relaxation rate ${1/T_1}$ was measured for $^{11}$B,  $^{195}$Pt and $^{7}$Li, and determined by a good fit of the recovery  of the nuclear magnetization to a single exponential function. 
 For the alloyed samples, $^{195}$Pt Knight shift was measured, since it is much larger than that of $^{11}$B or $^{7}$Li and provides higher accuracy   for  broadened spectra due to alloying.  Measurements below 1.4 K were carried out with a $^3$He-$^4$He dilution refrigerator. 

The electronic structure calculations were performed 
by using the full-potential augmented plane-wave plus local orbital method and GGA-PBE exchange-correlation function \cite{Perdew} as implemented in the WIEN2k code \cite{Blaha}. Spin-orbital interaction was included by using a second variational procedure. The muffin-tin radii were set to $R_{\rm MT}$=1.88 Bohr for B, $R_{\rm MT}$=2.27 Bohr for Li, and $R_{\rm MT}$=2.12 Bohr for Pd and Pt. The plane-wave cutt-off ($K_{\rm max}$) was determined by $R_{\rm min}$ $\cdot$ $K_{\rm max}$=7.0, where $R_{\rm min}$ is the minimal $R_{\rm MT}$. 


 \begin{figure}[htbp]
  \begin{center}
\includegraphics[width=90mm]{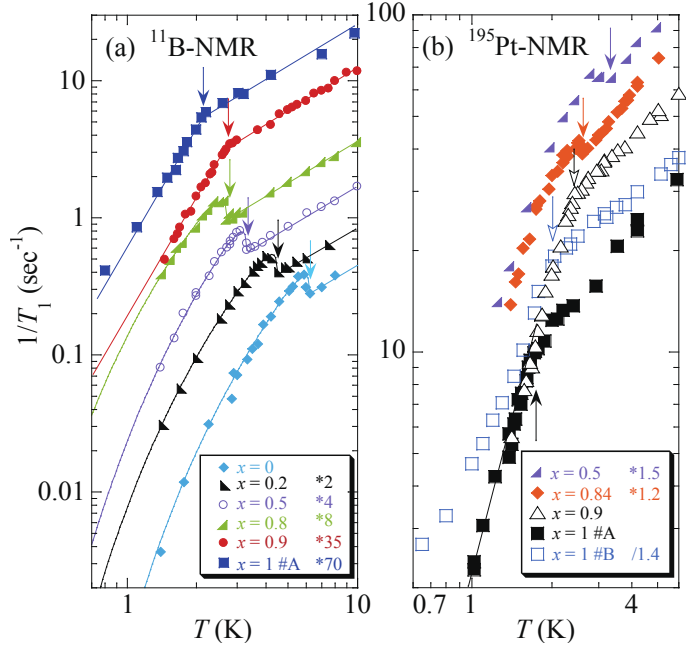}
 \end{center}
  \caption{
(Color online)  (a) The $T$ dependence of $^{11}(1/T_1)$.
 The straight lines above $T_c$ indicates the $1/T_1 \propto T$ relation.
(b) The $T$ dependence of $^{195}(1/T_1$)  for $x$ = 0.5, 0.84, 0.9 and 1.  For both (a) and (b), data were offset  vertically for clarity by multiplying or dividing by a number shown in the figure \cite{Supple}. Data for $x$=1 \#A ($H$=0.4 T) were taken from  \cite{nishiyama2}.The arrows indicate $T_{\rm c}$ under a magnetic field. The straight lines below $T_{\rm c}$ for  $x$=0.9 and 1  indicate the $1/T_1 \propto T^3$ relation. 
}
\label{T1-T(B)}
\end{figure}


Figure \ref{T1-T(B)} shows the temperature ($T$) dependence of $^{11}$B-NMR $1/T_1$ for various $x$.
For $x\leq$ 0.8, $^{11}(1/T_1)$ just below $T_{\rm c}$ is enhanced over its normal-state value, which is a well-known characteristic for an isotropic  energy gap. The data below $T_{\rm c}$   can be fitted by the BCS theory in a procedure described in  previous papers  \cite{nishiyama1,tahara}, with resulting gap amplitude  $\Delta_0$ = 1.70, 1.56, 1.75, 1.50 $k_{\rm B}T_{\rm c}$ for $x$ = 0, 0.2, 0.5, 0.8, respectively. The parameter $r=\Delta (0)/\delta$ that characterizes the height of the coherence peak is 1.8, where $\delta$ is the energy-level broadening. 
For $x$ = 0.9 and 1, however, $^{11}(1/T_1)$ shows no coherence peak just below $T_{\rm c}$ and is in proportion to $T^3$, which indicates the existence of line nodes in the  gap function. \par 


The contrasting behavior for the two groups  of $x\leq$0.8 and $x\geq$0.9 is seen in  $^{195}$Pt NMR as well.  In Fig. \ref{T1-T(B)} (b) is shown the $T$ dependence of $^{195}(1/T_1)$ for $x$ =0.5, 0.84,  0.9 and 1. For $x$ = 0.9 and 1,
$^{195}(1/T_1)$ shows no coherence peak  below $T_{\rm c}$ and decreases in proportion to $T^3$.  

\begin{figure}[htbp]
  \begin{center}
\includegraphics[width=60mm]{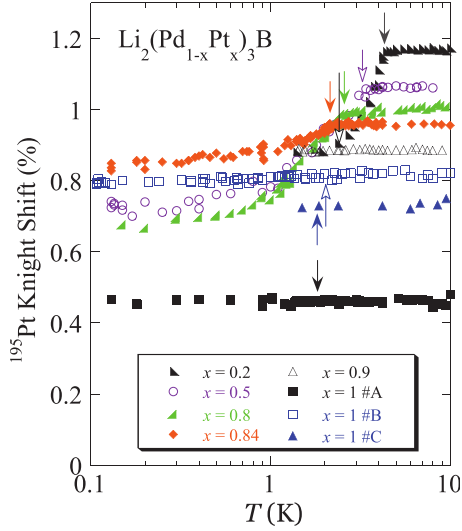}
 \end{center}
  \caption{
 (Color online) Temperature dependence of the $^{195}$Pt-Knight shift for various $x$. The arrows indicate $T_c$ at $H$=0.26 T except $x$=1 \#A. Data for $x$=1 \#A were taken from  \cite{nishiyama2}  ($H$=0.4 T) but re-calculated using the same magnetic-field calibration as other samples.
}
\label{K-T}
\end{figure}

To see  the spin state of the Cooper pairs, the spin susceptibility  $\chi_{\rm s}$ via  Knight shift measurement is the most effective probe.
The $T$ dependence of $^{195}$Pt Knight shift ($^{195}K$) for various $x$
is shown in Fig. \ref{K-T}. The $^{195}K$ for  $x$ = 0.2, 0.5 and 0.8 decreases below $T_{\rm c}$. So does it for $x$ =0.84,  but the reduction is smaller. For $x$ = 0.9 and 1, however, $^{195}K$  remains unchanged across $T_{\rm c}$. \par

\begin{figure}[htbp]
  \begin{center}
\includegraphics[width=55mm]{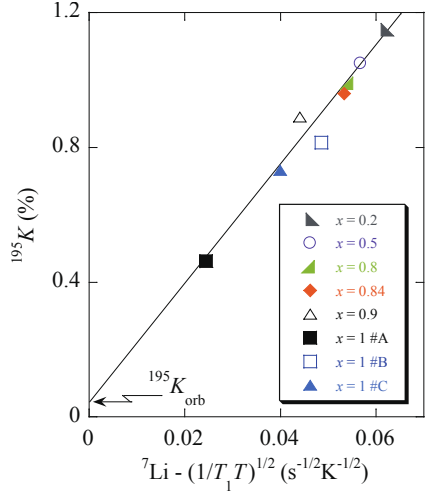}
 \end{center}
  \caption{ (Color online)  The $^{195}$Pt Knight shift in the normal state versus $^{7}(1/T_1T)^{1/2}$ for different Pt-contents.} 
\label{KsKn}
\end{figure}

In order to evaluate  quantitatively the evolution of $\chi_{\rm s}$,  one needs to know the Knight shift due to orbital susceptibility, $K_{{\rm orb}}$, since generally $K=K_{{\rm orb}}+K_{\rm {s}}$, where the spin part
$K_{\rm {s}}$ is proportional to $\chi_{\rm s}$ or the density of states (DOS) at the Fermi level, $N(E_F)$.  
$^7$Li NMR is useful for estimating $^{195}K_{\rm orb}$. All $^7$Li electrons are in $s$-orbits and the angular moment $L$ =0, 
so that $^7({1/T_1T}) \propto N(E_F)^2$ without orbital contribution mirrors $^{195}K_s$.
Figure \ref{KsKn} shows the relationship between $^{195}K$ and $^7(1/T_1T)^{1/2}$ for various Pt-contents. Indeed, a linear relation between the two quantities is seen.  From the extrapolation of the straight line, it is found that $^{195}K_{\rm orb}\approx$ 0.03\%,  then $^{195}K$  is due predominantly to $\chi_{\rm s}$. \par

These results  imply that the spin-triplet state is dominant for $x$ = 0.9 and 1 and it 
 evolves abruptly above $x=0.8$. 
In the intermediate regime, $x$=0.84, both the spin-singlet and spin-triplet characteristics can be seen.  
Namely, a small coherence peak  is observed and  the reduction of $^{195}K_{\rm s}$ below $T_c$ is much smaller.
Finally,  $^{195}K_{\rm s}$ not vanishing completely  for $x\leq$0.8  \cite{note0} may be understood as    due to some mixed spin-triplet component and the spin-flip scattering by disorder \cite{anderson,note1}, and even possible inter-band susceptibility \cite{note2}.  


\begin{figure}[htbp]
  \begin{center}
\includegraphics[width=70mm]{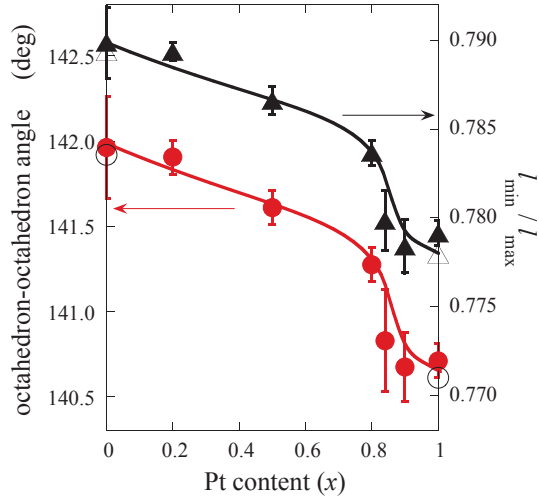}
\end{center}
  \caption{(Color online)  
 The $x$-dependence of octahedron-octahedron angle, $\alpha$, and the bond-length  ratio $\frac{l_{\rm min}}{l_{\rm max}}$. For $x$=1, data for sample \#B are shown. The open marks are data cited from Ref. \cite{eibenstein}. The curves are guides to the eyes.} 
\label{K-x-bond}
\end{figure} 

What is the origin of the abrupt increase of the spin triplet component above $x$=0.8? Below we show evidence for  the SOC increasing abruptly above $x$=0.8 due to an abrupt increase of the B(Pd,Pt)$_6$ octahedra distortion.
The extent of inversion-symmetry breaking can be measured by two parameters, namely, the angle  $\alpha$ between the two connecting octahedra (see Fig. 1(b)) and the ratio $\frac{l_{\rm min}}{l_{\rm max}}$, where $l_{\rm max}$ ($l_{\rm min}$) is the longest (shortest) M-M (M=Pd,Pt) bond length. In a centrosymmetric structure,  $\frac{l_{\rm min}}{l_{\rm max}}$=1 and $\alpha$=180$^{\circ}$. 
Figure \ref{K-x-bond} shows the $x$-dependencies of $\alpha$ determined by the general coordinates of Pd(Pt),   $12d$ $(\frac{1}{8}, y, \frac{1}{4}-y)$ and Li, and of $\frac{l_{\rm min}}{l_{\rm max}}$ determined by $y$ and the lattice constant, $a$ \cite{Supple}. 
Our data for $x$=0 and 1 are in good agreement with those calculated from $y$ and $a$ reported in \cite{eibenstein}. 
It is found that both $\alpha$ and $\frac{l_{min}}{l_{max}}$ show an abrupt reduction above $x$=0.8.

\begin{figure}[htbp]
  \begin{center}
\includegraphics[width=80mm]{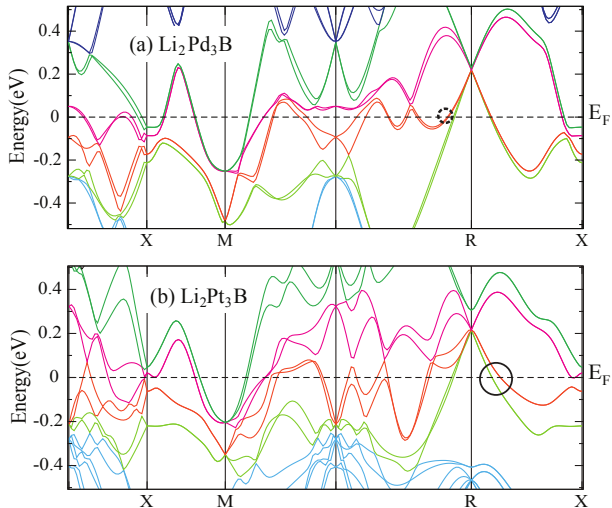}
\end{center}
  \caption{(Color online)  Calculated band dispersion for Li$_2$Pd$_3$B and Li$_2$Pt$_3$B. } 
\label{bandstructure}
\end{figure}

The band calculation indicates that such abrupt change in the local crystal structure as to increase the extent of the inversion-symmetry breaking enhances the SOC abruptly. Figure  \ref{bandstructure} shows the electronic band structure for Li$_{2}$Pd$_{3}$B and Li$_{2}$Pt$_{3}$B, using the crystal structure data obtained in this work \cite{Supple}, which  is in agreement with that reported previously \cite{Pickett}. 
As shown in Fig. 6, the energy bands with SOC are quite complex. Both the splitting due to SOC near $E_F$ around the R-point and X-point are strong. However, only one of SOC-split  bands around X-point crosses $E_F$ while the two of SOC-split bands around R-point cross $E_F$. Thus, in order to simply analyze the effect of SOC-splitting, below we choose the SOC-splitting around R as a typical one.
%

\begin{figure}[htbp]
  \begin{center}
\includegraphics[width=55mm]{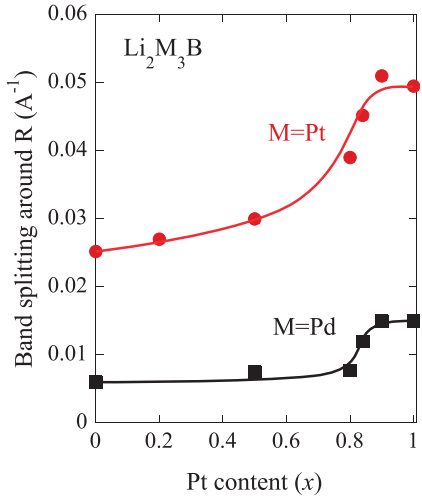}
\end{center}
  \caption{(Color online)  Band splitting around the R point assuming Pt (Pd) occupies all the M sites. The horizontal axis $x$ means  that the crystal structure data for Li$_2$(Pd$_{1-x}$Pt$_x$)$_3$B are used in the calculation.  The curves are guides to the eyes. } 
\label{band-splitting}
\end{figure}

We iterated the calculation for all composition $x$  using the lattice parameter and the ionic positions for each $x$ obtained from the Rietvelt analysis but assuming  that all M sites are occupied by Pt. 
The resulting band splitting around the R point along the R-X direction, as marked by the circle in Fig. \ref{bandstructure} (b),  is shown in Fig. \ref{band-splitting} by solid circles. 
Then we performed the same calculation assuming that all M sites are occupied by Pd. In this case, it is the  bands  around the R point but along the R-$\Gamma$ direction, as marked by the broken circle in Fig. \ref{bandstructure} (a), that show a pronounced $x$-dependence (squares in Fig. \ref{band-splitting}). In both cases, the splitting increases abruptly at $x$=0.8. Also,  the eight-degenerate band at the R point around $E$=0.2 eV is split into a two-fold band and a six-fold band, forming a gap between them \cite{Supple}. That  gap  also shows an abrupt increase at $x$=0.8 \cite{Supple}. 
This result clearly indicates  that, although Pt plays an important role as seen by the difference between the data shown by circles and squares in Fig. \ref{band-splitting}, the distortion of the B(Pt,Pd)$_6$     octahedron is more effective in enhancing the SOC which is  called the asymmetric SOC in some literatures. Since  the mixing of the spin-singlet and spin-triplet states is determined by the strength of the SOC \cite{gorkov, frigeri1}, the experimental results can be understood as that the asymmetric SOC for $x\geq$0.84  is so large that the spin-triplet state with line nodes becomes dominant. Although the recently discovered NCS superconductors contain heavy elements such as Pt, Re or Ir \cite{klimczuk,tahara,Nohara,Wen,Bauer2,Eguchi},  they lack the structural element found here for $x\geq$0.84. This appears to be the reason why those NCS materials exhibit conventional superconducting properties. Finally, previous measurements on Li$_2$(Pd$_{0.5}$Pt$_{0.5}$)$_3$B by $\mu$sr \cite{Hafliger} and upper critical field \cite{Peets} highlighting the $s$-wave gap characteristics are consistent with the present  results. On the other hand, some extent of mixing of the spin-triplet component was suggested by analyzing the specific heat \cite{takeya} and penetration depth \cite{yuan2} data on Li$_2$(Pd$_{0.5}$Pt$_{0.5}$)$_3$B. In this regard, theoretical and experimental means  to estimate quantitatively the mixed spin-triplet component from the NMR  data are desired.

In conclusion, we have  
studied the evolution of the superconducting state in  Li$_2$(Pd$_{1-x}$Pt$_x$)$_3$B by  NMR. 			
We find that the pairing symmetry changes drastically at $x$ = 0.8. For $x \leq$ 0.8, the materials are in a predominantly spin-singlet   state. 
However, for $x>$0.8, unconventional properties  due to the mixing of  the spin-triplet state appear. 
The change is caused by an abrupt enhancement of the asymmetric SOC 
due to an increased distortion of the B(Pd,Pt)$_6$ octahedral units.
Our results indicate that, in addition to a large $Z$,  the structure distortion as to increase the extent of inversion-symmetry breaking is another important, and more effective factor to increase  the mixing of the spin-triplet state.


We thank  G.Z. Bao, K. Arima,  S. Kawasaki, and Y. Takabayashi for help in experiments and analysis, and K. Miyake, Y. Fuseya, T. Shishido, K. Asayama, and K. Matano  for helpful discussions.
Work in Okayama was supported by MEXT grant 
No. 22103004 and JSPS grant No.20244058. Work in Beijing was supported by NSFC (Grants No. 10974231 and
No. 11174337) and MOST of China (Grants No. 2011CBA00100 and No. 2011CBA00109).


\begin{thebibliography}{00}
\bibitem{note0}
It should be emphasized that the result is not due to non-superconducting component of the sample, since the nuclear magnetization recovery is of single component and the spectra do not show two-component characteristics.
\bibitem{note1}
Indeed, the NMR FWHM   increases to 12.3 kHz ($x$=0.8), 12.5 kHz ($x$=0.5) and 13.1 kHz ($x$=0.2) from 8.5 kHz for ($x$=0.9, $x$=1 \#A) and 11 kHz for $x$=0.84, which suggests that  spin-flip scattering can be larger for  0.2$\leq x \leq$0.8.
\bibitem{note2}
Such contribution $\chi\propto \sum_{k_+, k_{-}}\frac{f({k_-})-f({k_+})}{E({k_+})-E({k_-})}$, where + and - denote the two split bands,  is  $x$-insensitive, if any. This is because, qualitatively speaking,  the   energy splitting $E({k_+})-E({k_-})$ and the population imbalance $f({k_-})-f({k_+})$ are both proportional to the SOC strength.
\end{thebibliography}
\end{document}